\documentclass[pra,showpacs,letterpaper,twocolumn,showpacs,superscriptaddress]{revtex4}
\usepackage{graphicx,psfrag,amsmath,amssymb,amsfonts,bbm,latexsym,color,dcolumn,bm}

\begin{document}

\title{Dynamics of Open Bosonic Quantum Systems in Coherent State Representation}

\author{D.A.R. Dalvit}
\affiliation{Theoretical Division, MS B213, Los Alamos National
Laboratory, Los Alamos, NM 87545, USA}

\author{G.P. Berman}
\affiliation{Theoretical Division, MS B213, Los Alamos National
Laboratory, Los Alamos, NM 87545, USA}

\author{M. Vishik}
\affiliation{Department of Mathematics, The University of Texas at Austin, Austin, TX 78712-1082, USA}

\date{\today}


\begin{abstract}

We consider the problem of decoherence and relaxation of open bosonic
quantum systems from a perspective alternative to the
standard master equation or quantum trajectories approaches. Our
method is based on the dynamics of expectation values of
observables evaluated in a coherent state representation. We
examine a model of a quantum nonlinear oscillator with a density-density
interaction with a collection of environmental oscillators at
finite temperature. We derive the exact solution for dynamics of
observables and demonstrate a consistent perturbation approach.

\end{abstract}

\maketitle

\section{Introduction}

When a small quantum system interacts with an environment it
undergoes irreversible processes such as relaxation and
decoherence. Environment-induced decoherence is at the heart of
the quantum-classical transition \cite{zurekRMP,joos}. Classicality is an emergent
property induced on subsystems by their environment. Under a variety of conditions, 
which are particularly easy to satisfy for macroscopic objects, decoherence leads to the selection
of a small subset of quasi-classical states from within the huge Hilbert space.

The dynamics of decoherence and relaxation in open quantum systems is
usually studied with the help of the master equation for the
reduced density matrix $\rho$ of the system \cite{leshouches}, or by means of any of its
possible unravellings \cite{unravellings}, including continuous quantum measurement and
quantum trajectories \cite{dalvitprl,dalvitpra1,dalvitpra2,wiseman,habib}. 
It is also possible to study the decoherence process in terms of phase space densities, especially used in the
field of quantum optics \cite{gardiner}.  In this context one expresses the density operator $\rho$ of a harmonic
oscillator (one mode of the electromagnetic field) in terms of a c-number function of a coherent state (complex) variable
$\alpha$. Particularly useful distributions are: a) The Q-function, $Q(\alpha,\alpha^*)=\frac{1}{\pi} \langle \alpha | \rho | \alpha \rangle$, 
that allows the calculation of antinormally ordered quantum expectation values in terms of simple moments of $Q(\alpha,\alpha^*)$; 
b) The P-function, $P(\alpha,\alpha^*)$,  defined  via $\rho=\int d^2 \alpha P(\alpha,\alpha^*) |\alpha \rangle \langle \alpha|$, 
whose moments give normally ordered quantum expectation values; and c) The Wigner function $W(\alpha,\alpha^*)$, whose moments
are equal to the expectation values of symmetrically ordered products of creation and annihilation operators. These phase-space 
distributions have certain drawbacks. For example, they may no be positive definite, or may make no sense for certain density matrices.
It may be difficult to extract physical information from these quasi-probability distributions, especially in the context of nonlinear open quantum
systems. 

The aim of this paper is to introduce an alternative approach based on c-number dynamical equations for
expectation values of observables of open bosonic quantum systems. The method is a
generalization of the well studied asymptotic theory for bosonic
closed quantum systems \cite{bermanbook} to the case of open dynamics.
Such an approach provides a method to derive exact, c-number, partial differential equations describing
the evolution of quantum averages evaluated in  coherent states.  In this sense the method is related to the phase
space distributions discussed in the above paragraph, but has the key feature of dealing directly with expectation
values rather than with the quantum state $\rho$. This change of focus has several advantages. First, the physical interpretation
of the results is clearer, and one does not need to compute integrals over moments to obtain physical quantities, which is
especially difficult in the quasi-classical regime of parameters due to fast oscillations. Second, 
the differential equations are well behaved in the quasi-classical limit $\hbar/ N \rightarrow 0$,
where $N$ is a quasi-classical parameter of the system, and lead to asymptotic Laplace-type expansions \cite{vishik}. 
The crucial property of Laplace asymptotics is that observable quantities are exponentially localized in phase space around coherent states, and
do not have the standard oscillatory WKB behavior \cite{maslov}. Third, it provides a simpler interpretation of singularities (in the sense of perturbation theory
of partial differential equations \cite{bermanbook}) that appear in the quasi-classical regime for nonlinear Hamiltonians, allowing one to distinguish which part
of the singularity is connected with measurable physical phenomena, and which is connected just with the choice of representation.

As a prototypical system we consider the dynamics of a quantum nonlinear oscillator (QNO)
\begin{equation}
\hat H_S = \hbar \omega \hat{a}^{\dagger} \hat{a} + \mu \hbar^2 ( \hat{a}^{\dagger} \hat{a})^2 ,
\label{qno}
\end{equation}
interacting with a bath of linear oscillators initially in thermal
equilibrium. Here $\hat{a}$ ($\hat{a}^{\dagger}$) are annihilation
(creation) bosonic operators, $\omega$ is the linear frequency and
$\mu$ is the parameter of nonlinearity. The QNO is initially
prepared in a coherent state $|\alpha\rangle$ in the
quasi-classical region of parameters. The quasi-classical
parameter is $\hbar / J \ll 1$, where $J=\hbar
|\alpha|^2$ is the action of the linear classical oscillator. 
The nonlinear harmonic oscillator Hamiltonian Eq. (\ref{qno}) may describe
a Bose-Einstein condensate (BEC) treated in the single mode approximation. 
Such an approximation is valid when the many-body interactions within the condensate
produce a small modification of the ground state of the trap, the mode structure being sparce,
such as in tightly optically trapped systems.  For BECs trapped in optical  lattices, the single-mode 
approximation leads to predictions for the quantum dynamics of the condensate in excellent agreement
with experiments \cite{greiner, cataliotti}. Other systems that can be described by this Hamiltonian
are  micro- \cite{chan} and nanomechanical \cite{cleland} resonators in the nonlinear regime,
and nonlinear optical systems, among others. 

The quantum and the classical dynamics of the anharmonic oscillator was studied
in detail in \cite{milburn86} using the Q phase-space distribution of the system. It was shown that the presence
of non-positive-definite second-order terms in the quantum evolution of Q, not present in the evolution of the 
classical probability distribution, is responsible for quantum recurrences and prevents the appearance of fine-scale-structure
``whorls" predicted in the classical description. In \cite{milburnholmes,milburndaniel}  the interaction of the nonlinear oscillator
with an environment (modeled by a thermal bath of harmonic oscillators with position-position coupling to the system oscillator)
was studied in the limit of small nonlinearity using the Q function formalism, and it was argued that such an interaction was effective in destroying quantum
interference effects and restoring the classical phase-space structure.  However, as recently shown by some of us in \cite{paperpra} by means
of solving the master equation for the reduced density matrix of the system,
environment-induced decoherence is in fact ineffective in recovering the quantum-classical correspondence for this nonlinear
system: Some quantum effects may survive the decoherence process, and be
observed for times much larger than the decoherence time-scale. In
particular, we showed that the Ehrenfest time $t_E = (2 \hbar \mu
|\alpha|)^{-1}$, which characterizes the departure of quantum
dynamics for observables from the corresponding classical
dynamics, can be observed for times much longer than the
decoherence time-scale.

The paper is organized as follows: In Section II we consider a model of phase decoherence that is exactly solvable, and will be used to exemplify our
method based on the coherent state representation for observable values of open bosonic quantum systems, that we describe in Section III. This section
contains the main results of this paper: The general theory proposed in this paper is first described, and it is then applied to the case of the nonlinear oscillator, presenting both exact and perturbative treatments. Finally, Section IV contains our conclusions and briefly discusses possible extensions of this work.


\section{Model for phase decoherence in the nonlinear oscillator}

In this section we consider the model of phase decoherence in the nonlinear oscillator, that allows for an exact solution for the reduced
dynamics of the system. This model will serve us for ease of presentation of the general theory of dynamics of open bosonic quantum systems
in coherent state representation that will be described in the next Section. The interaction between the nonlinear oscillator and the thermal bath
of harmonic oscillators with Hamiltonian
\begin{equation}
\hat{H}_E =\sum_{j=1}^N \hbar \omega_j \hat{b}_j^{\dagger} \hat{b}_j ,
\end{equation}
is of the density-density type 
\begin{equation}
\hat{H}_{\rm int} = \sum_j^N  g_j  \hat{a}^{\dagger} \hat{a} \hat{b}_j^{\dagger} \hat{b}_j .
\end{equation}
Here $\hat{b}_j$ ($\hat{b}_j^{\dagger}$) are annihilation/creation operators of the
environment, $\omega_j$ are the frequencies of the environmental
harmonic oscillators, $g_j$ are coupling constants, and $N$
is the total number of the oscillators in the environment. This model
of decoherence can be used as an approximate description for the collisional
effects between an ultracold atomic gas and its thermal cloud \cite{decobec,savage}. 
Indeed, for low enough temperatures phase decoherence (corresponding to density-density type of interactions, 
{\it i.e.}, elastic two-body collisions that conserve the total number of condensed particles) dominate over amplitude
decoherence processes (that correspond to position-position coupling, {\it i.e.}, inelastic two-body processes that either
feed or deplete the condensate).  

The joint system-environment dynamics can be exactly solved
in the Fock basis. Let us assume that the initial joint state is
uncorrelated, {\it i.e.}, $\rho_{\rm tot}(0)= \rho(0) \otimes \rho_E(0)$.
The initial state of the system is assumed to be a pure coherent state
$\rho(0) = | \alpha \rangle \langle \alpha|$, and the initial state of the environment is assumed to be
a thermal state $\rho_E = Z^{-1}_E e^{- \hat{H}_E / k_{\rm B} T}$, where
$Z_E = {\rm Tr}_E [ e^{- \hat{H}_E / k_{\rm B} T}] $ is the partition function of the environment. 
Here $T$ is the temperature of the bath, and $k_{\rm B}$ is Botzmann constant.  The joint density matrix at time $t$ can be
easily computed in the number representation since both the system, the environment, and the interaction Hamiltonians are diagonal
in that basis. The result is 
\begin{eqnarray}
&&\rho_{\rm tot}(t) = \sum_{n,n'=0}^{\infty} a_n a_{n'}^* e^{-i t [ \omega (n-n') + \mu \hbar (n^2-n'^2)]} |n \rangle
\langle n' | \otimes \nonumber \\
&& Z_E^{-1} \prod_{j=1}^N \sum_{\mu^{(j)}=0}^{\infty} e^{- \frac{\hbar \omega_j \mu^{(j)}}{ k_{\rm B} T}}
e^{- \frac{i g_j t}{\hbar} (n-n') \mu^{(j)}} | \mu^{(j)} \rangle \langle \mu^{(j)} | . \nonumber
\end{eqnarray}
Here $a_n= e^{-|\alpha|^2/2}  \alpha^n/ \sqrt{n!}$ are the expansion coefficients of the coherent
state $|\alpha \rangle$ in the Fock basis.

Tracing over the environmental degrees of freedom it is easy to find the reduced density
matrix for the nonlinear oscillator $\rho(t)$, and calculate any expectation value of a  system operator.
In particular, the evolution of the coherent state amplitude  is given by
\begin{equation}
\langle \hat{a}(t) \rangle = \alpha(t) R(t),
\label{me1}
\end{equation}
where $\alpha(t)$ is the solution without coupling to the environment, namely
\begin{equation}
\alpha(t)  = \alpha \; e^{-i (\omega+ \mu \hbar ) t}
\; \exp[ |\alpha|^2 (e^{-2 i \mu \hbar t}-1) ] ,
\label{me2}
\end{equation}
and $R(t) =  \prod_j R_j(t)$ is the decoherence factor, that contains the effects due to the environment. Each $R_j$ can be written
in terms of its modulus and phase, $R_j(t) = | R_j(t)| e^{i \varphi_j(t)}$, where
\begin{eqnarray}
| R_j(t) | &=& \frac{1- e^{- \frac{\hbar \omega_j}{k_{\rm B} T}}}
{\sqrt{1-2 e^{- \frac{\hbar \omega_j}{k_{\rm B} T}} \cos(g_j t/\hbar) + e^{- \frac{2 \hbar \omega_j}{k_{\rm B} T}}}} , \nonumber \\
\tan \varphi_j(t) &=& - \frac{e^{- \frac{\hbar \omega_j}{k_{\rm B} T}} \sin(g_j t/\hbar)}{1-e^{- \frac{\hbar \omega_j}{k_{\rm B} T}} \cos(g_j t/\hbar) } .
\label{eres}
\end{eqnarray}
The coherent amplitude $\alpha(t)$  departs from the classical solution
on the Ehrenfest time scale $t_E = 1/2 \hbar \mu |\alpha|$, and undergoes collapses and revivals,
the revival time being $t_R = \pi/ \hbar \mu$ \cite{paperpra}. The effect of the environment is to produce a dephasing
of the coherent state amplitude $\alpha(t)$, causing it to decay to zero in the limit of a large
environment ($N \gg 1$), and killing the revivals. For finite $N$, the decay is incomplete and the
revivals are suppressed. In the special case of identical environmental oscillators ($g_j \equiv g$ and
$\omega_j \equiv \omega_E$ for all $j$), the decoherence factor $R(t)$ can be approximated in the
limit of large $N$  by a periodic Gaussian structure. Each Gaussian $R_p(t)$ is centered around
a time $t_p = 2 \pi \hbar p / g$ ($p=0,1, 2, \ldots$) and has the form
\begin{equation}
R_p(t) = \exp \left[ - \frac{g^2 N (t-t_p)^2}{2 \hbar^2} \; \frac{e^{-\frac{\hbar \omega_E}{k_{\rm B} T}}}{(1-e^{- \frac{\hbar \omega_E}{k_{\rm B} T}})^2} \right] .
\nonumber
\end{equation}


\section{Coherent state representation}

There are several ways of considering the dynamical behavior of quantum expectation values. One possibility is to solve the Heisenberg equation for the
density matrix in some basis states and then take expectation values of the corresponding operators, as done in the previous Section. For general nonlinear
Hamiltonians this approach leads, in the quasiclassical asymptotic limit, to a singular behavior (in the sense of perturbation theory of partial differential equation)
\cite{bermanbook}. Instead, it is possible to write down exact, c-number partial differential equations for expectation values, which allows one to distinguish
which part of this singularity is connected with measurable physical phenomena, and which is connected just with the choice of representation.  This method is
described extensively in \cite{bermanbook} for closed bosonic quantum systems, and it is based on computing quantum observables in the coherent state 
basis. In the following we briefly review the methodology for this case, and then we generalize it to open systems, {\it i.e.}, quantum systems in interaction with
an external environment.


\subsection{Closed systems}

In this subsection we exemplify the methodology of coherent state representation for closed systems using the nonlinear oscillator described with the Hamiltonian
Eq.(\ref{qno}). Similar  ideas can be applied to any quantum boson and spin systems, as described in \cite{bermanbook}. 
Given an arbitrary operator of the system $\hat{f}_S = \hat{f}_S(\hat{a}^{\dagger}, \hat{a})$, it is possible to write down an exact, c-number partial differential
equation for the time-dependent expectation value $f_S(\alpha^*,\alpha,t) = \langle \alpha| e^{i \hat{H}_S t/\hbar} \hat{f}_S e^{-i \hat{H}_S t/\hbar} |\alpha \rangle$
of such operator evaluated in coherent states $| \alpha \rangle$.  Using Heisenberg equation $d/dt \hat{f}_S = (i/\hbar) [ \hat{H}_S, \hat{f}_S ] $, it follows
\begin{equation}
\dot{f}_S = \frac{i}{\hbar} \; ( \langle \alpha | \hat{H}_S \; \hat{f}_S | \alpha \rangle - \langle \alpha | \hat{f}_S \;  \hat{H}_S | \alpha \rangle ) .
\label{fcoh}
\end{equation}
Putting both $\hat{H_S}$ and $\hat{f}_S$ into normal-ordering form in terms of the initial operators 
$\hat{a} \equiv \hat{a}(t=0)$ and $\hat{a}^{\dagger} \equiv  \hat{a}^{\dagger}(t=0)$, it is possible to obtain a closed form for each of the two terms in Eq.(\ref{fcoh}). 
One obtains
\begin{eqnarray}
\langle \alpha | \hat{H}_S \; \hat{f}_S | \alpha \rangle &=& e^{- |\alpha|^2} \; H_S \left( \alpha^*, \frac{\partial}{\partial \alpha^*} \right) \; f_S \; e^{|\alpha|^2} ; \nonumber \\
\langle \alpha | \hat{f}_S \;  \hat{H}_S | \alpha \rangle &=& e^{- |\alpha|^2} \; H_S \left( \alpha, \frac{\partial}{\partial \alpha} \right) \; f_S \; e^{|\alpha|^2} .
\end{eqnarray}
Here $f_S=f_S(t)$ is the c-number expectation value we seek, and the differential operator $H_S$ has the same functional form of the normal-ordered
operator function $\hat{H}_S(\hat{a}^{\dagger}, \hat{a} )$, but with the substitution $\hat{a}^{\dagger} \rightarrow \alpha^*$, and 
$\hat{a} \rightarrow \frac{\partial}{\partial \alpha^*}$. Therefore, the exact partial differential equation for the time-dependent expectation value $f_S(t)$ reads
\begin{equation}
\frac{\partial f_S}{\partial t} = \hat{K}_S  \; f_S , 
\end{equation}
with initial value $f_S(0)=f_S(\alpha^*, \alpha)$, and the partial differential operator $\hat{K}_S$ given by
\begin{equation}
\hat{K}_S = \frac{i}{\hbar} e^{-|\alpha|^2} \left[ H_S \left(\alpha^*, \frac{\partial}{\partial \alpha^*} \right) - H_S \left( \alpha,\frac{\partial}{\partial \alpha} \right) \right] \; 
e^{|\alpha|^2} .
\end{equation}
This operator can be split into two parts: $\hat{K}_S = \hat{K}_{\rm cl} + \hbar \hat{K}_q $.
The first operator $\hat{K}_{\rm cl}$ includes only first order derivatives and describes the corresponding classical limit, while the second operator $\hat{K}_q$ includes higher order derivatives and is responsible for quantum effects. For the model described by Eq. (\ref{qno}) the exact partial differential equation
for observables  reads
\begin{eqnarray}
\frac{\partial f_S}{\partial t} &=& i (\omega + \hbar \mu + 2 \hbar \mu |\alpha|^2) \left(
\alpha^* \frac{\partial}{\partial \alpha^*} - \alpha \frac{\partial}{\partial \alpha} \right) f_S + \nonumber\\
&& i \hbar \mu \left( (\alpha^*)^2 \frac{\partial^2}{\partial (\alpha^*)^2} -
\alpha^2 \frac{\partial^2}{\partial \alpha^2} \right) f_S.
\label{noenv}
\end{eqnarray}
In particular, for $\hat{f}_S = \hat{a}$, the evolution of $f_S(t)$ corresponds to the evolution of the coherent
amplitude $\alpha(t) = \langle \alpha| \hat{a}(t) | \alpha \rangle$, and the solution is the same as in
Eq. (\ref{me2}).


\subsection{Open systems: Exact treatment}

In order to extend the formalism to treat open quantum systems, we
assume that the system, initially populated in a coherent state $|
\alpha \rangle$, interacts with the bath of harmonic oscillators,
also initially populated in coherent states $\{ | \beta_j \rangle
\}$.  Let $\hat f$ be any operator of the composite system
\begin{equation}
\hat{f} = \hat{f} ( \hat{a}^{\dagger}(t),\hat{a}(t), \{ \hat{b}_j^{\dagger}(t) \}, \{ \hat{b}_j(t) \}) ,
\end{equation}
that evolves according to the Heisenberg  equation $d \hat{f} / dt  =
\frac{i}{\hbar} \; [ \hat{H},  \hat{f} ]$, where $\hat{H} =
\hat{H}_S + \hat{H}_E + \hat{H}_{\rm int}$ is the Hamiltonian of
the composite system. The initial state is $| \hat{\Psi}(0)
\rangle = |\alpha, \{ \beta_j \} \rangle$. The expectation value
$f(t)$ of the operator $\hat f(t)$ in the state $|
\hat{\Psi}(0)\rangle $, evolves according to the partial differential
equation 
\begin{equation}
\frac{ \partial f}{\partial t}  = \hat{K} f , 
\end{equation}
where $\hat{K}$ is the differential operator
\begin{eqnarray}
\hat{K} &=& \frac{i}{\hbar} e^{ - |\alpha|^2 - \sum_j |\beta_j|^2 }
\left[
H
\left(
\alpha^*, \{ \beta_j^* \}, \frac{\partial}{\partial \alpha^*}, \left\{ \frac{\partial}{\partial \beta_j^*} \right\}
\right)  -   \right. \nonumber \\
&& \left.
H
\left(
 \alpha, \{ \beta_j \} ,
\frac{\partial}{ \partial \alpha}, \left\{\frac{\partial}{\partial \beta_j} \right\} \right)
\right]
e^{  |\alpha|^2 + \sum_j |\beta_j|^2 }  .
\end{eqnarray}
Note that the following substitutions have been used: $\hat{a}^{\dagger} \rightarrow \alpha^*$, $\hat{a} \rightarrow \frac{\partial}{\partial \alpha^*}$, $\hat{b}^{\dagger}_j \rightarrow \beta_j^*$,
and $\hat{b}_j \rightarrow \frac{\partial}{\partial \beta_j^*}$.
For the model under consideration, the c-number differential equation for the composite (system+bath) expectation value $f(t)$ has the form
\begin{equation}
\frac{ \partial f}{  \partial t } = (\hat{K}_{\alpha} + \hat{K}_{\beta} + \hat{K}_{\rm int}) f .
\label{diff}
\end{equation}
The first two terms in Eq. (\ref{diff}) correspond to the free system-bath dynamics,
\begin{eqnarray}
\hat{K}_{\alpha} f  &=& i (\omega + \mu \hbar + 2 \mu \hbar | \alpha|^2)
\left( \alpha^* \frac{\partial}{\partial \alpha^*} - \alpha \frac{\partial}{\partial \alpha} \right) f  + \nonumber \\
&& i \mu \hbar \left[ ( \alpha^*)^2 \frac{\partial^2}{\partial(\alpha^*)^2} -
\alpha^2 \frac{\partial^2}{\partial \alpha^2} \right] f , \\
\hat{K}_{\beta} f  &=&  i \sum_j \omega_j \left( \beta^*_j \frac{\partial}{\partial \beta^*_j} -
 \beta_j \frac{\partial}{\partial \beta_j} \right) f ,
\end{eqnarray}
and the last term in Eq. (\ref{diff}) is due to the system-bath
interaction. It can be written as the sum of three contributions
$\hat{K}_{\rm int} f = (\hat{K}_{\rm int}^{(1)}  + \hat{K}_{\rm
int}^{(2)}  + \hat{K}_{\rm int}^{(3)} ) f $, where
\begin{eqnarray}
\hat{K}_{\rm int}^{(1)} f &=& \frac{i}{\hbar}  \left( \sum_j g_j |\beta_j|^2 \right) \;
\left( \alpha^* \frac{\partial}{\partial \alpha^*} - \alpha \frac{\partial}{\partial \alpha} \right) f, \nonumber \\
\hat{K}_{\rm int}^{(2)} f &=& \frac{i}{\hbar} | \alpha|^2 \sum_j g_j \left( \beta^*_j \frac{\partial}{\partial \beta^*_j} -
\beta_j \frac{\partial}{\partial \beta_j} \right) f,  \nonumber \\
\hat{K}_{\rm int}^{(3)} f &=& \frac{i}{\hbar} \sum_j g_j \left( \alpha^* \frac{\partial}{\partial
\alpha^*} \beta^*_j \frac{\partial}{\partial \beta^*_j} -
\alpha \frac{\partial}{\partial \alpha} \beta_j \frac{\partial}{\partial \beta_j} \right) f . \nonumber
\end{eqnarray}
Given a solution  $f(\alpha,\alpha^*, \{ \beta_j \}, \{ \beta^*_j \};t)$ to Eq. (\ref{diff}), we  finally have to
trace over the coherent states $\{ \beta_j \}, \{ \beta^*_j \}$ ({\it i.e.}, trace over the environment)
to obtain the evolution of expectation values of the system.

In the above we have assumed that initially each j-th environmental oscillator is in a pure coherent state $| \beta_j \rangle$.  Let us now consider
the case in which each environmental oscillator is initially in a mixed, thermal state at temperature $T$. Since the oscillators in the environment are
non-interacting, the initial density matrix of the environment can be written as a direct product over individual density matrices for each j-th sub-environment,
that is $\rho_E(0) = \prod_{j=1}^N \rho_E^{(j)}(0)$. The density matrix of each environmental oscillator is then
$\rho_E^{(j)}(0)= (Z_E^{(j)})^{-1} \; \exp(- \hbar \omega_j b_j^{\dagger} b_j / k_B T)$, where $Z_E^{(j)} = {\rm Tr}[ e^{- \hbar \omega_j b_j^{\dagger} b_j / k_B T} ]$
is the partition function of the j-th environmental oscillator. This mixed thermal state $\rho_E^{(j)}(0)$ can be written in the coherent state basis 
(this corresponds to the so-called P-representation or coherent state representation \cite{scully})
\begin{equation}
\rho_E^{(j)}(0) = \int d^2 \beta_j \; P(\beta_j^*, \beta_j) \; |\beta_j \rangle \langle \beta_j | ,
\end{equation}
where the probability distribution for each $\beta_j$ is given by 
\begin{equation}
P(\beta^*_j, \beta_j ) = \frac{1}{\pi \bar{n}_j}  \; e^{- |\beta_j|^2 / \bar{n}_j} .
\end{equation}
Here $\bar{n}_j = (e^{\hbar \omega_j/ k_{\rm B} T} - 1)^{-1}$ is the Bose distribution. That is, the coherent state representation
of a thermal state has a Gaussian distribution.

For the particular case of the system oscillator initially prepared in a coherent state, we would
like the solution to Eq. (\ref{diff}) to lead us to Eq. (\ref{me1}), once the integration over the
environmental variables is performed. Since the structure of Eq. (\ref{me1}) is the product
of the free evolution solution $\alpha(t)$ times a time-dependent factor that arises from
the coupling with the environment, we propose a solution to
Eq.(\ref{diff}) of the form
\begin{equation}
f(\alpha,\alpha^*, \{ \beta_j \}, \{ \beta^*_j \} ;t) = f_{\alpha} (\alpha, \alpha^*;t) \times
f_{\beta}(\{ \beta_j \}, \{ \beta_j^* \}; t) ,
\end{equation}
where $f_{\alpha}=\alpha(t)$ is given by Eq. (\ref{me2}).
The reduced dynamics of an observable of the system will be given by an average over the
environmental oscillators weighted by their respective probability distributions
\begin{equation}
f_S(t)  = \alpha(t) \times \int \prod_j d^2 \beta_j P(\beta^*_j,\beta_j) f_{\beta}(t) .
\label{reduced}
\end{equation}
The initial condition for $f_{\beta}(t)$ is $f_{\beta}(t=0)=1$. Since $P(\beta^*_j,\beta_j)$ depends on $\beta_j$ through its modulus squared, we can assume, without loss of generality, that at any
time $t$ the function $f_{\beta}(t)$ depends on $|\beta_j|^2$ (any other dependency,
like $\beta_j^k (\beta_j^{*})^m$ ($k \neq m$), vanishes upon integration).
This implies that the operator $\hat{K}_{\beta}$ acts on $f_{\beta}$ as
$\hat{K}_{\beta} f_{\beta}= 0$.
Given that $\dot{f}_{\alpha} = \hat{K}_{\alpha} f_{\alpha}$, the
equation for $f_{\beta}$ finally reads
\begin{equation}
\dot{f}_{\beta} = f_{\alpha}^{-1} \hat{K}_{\rm int} f_{\alpha} f_{\beta} .
\end{equation}
As can be shown by direct inspection, an exact solution to this equation, with initial
condition $f_{\beta}(t=0)=1$,  is
\begin{equation}
f_{\beta}(t) = \prod_j f_{\beta}^{(j)}(t) = \prod_j \exp \left[ - |\beta_j|^2 (1-e^{-i g_j t/\hbar}) \right] .
\end{equation}
Integrating upon the probability distributions $P(|\beta_j|^2)$ we obtain
the reduced dynamics for the observable of the system oscillator, namely
\begin{equation}
f_S(t) = \alpha(t) \prod_j \int d^2 \beta_j \; P(|\beta_j|^2) \;
f_{\beta}^{(j)} = \alpha(t) R(t),
\end{equation}
where $R(t) = \prod_j R_j(t)$ is given in Eq. (\ref{eres}). This expression
coincides with the exact solution derived in Eq. (\ref{me1}), obtained from
solving the Heisenberg equation for the density matrix of the joint quantum nonlinear oscillator-environment system. Therefore, using our method based on the coherent state representation for observables values for open bosonic quantum systems, we can recover the exact reduced dynamics.

\subsection{Open systems: Perturbative treatment}

There are a few models of decoherence and relaxation for which it is possible to solve exactly the joint system-environment dynamics, and to write
down and solve an exact master equation for the reduced density matrix of the system. Examples are the one considered in Section II, that can be trivially
solved in the number representation, and  the well-known quantum Brownian motion model, in which a linear oscillator ($\mu=0$ in Eq.(\ref{qno})) is
coupled through position to a bath of linear harmonic oscillators. This latter model can be solved using, for example, influence functional techniques
\cite{qbm} thanks to the fact that both the system, the bath, and the interaction Hamiltonians are quadratic forms. For other general models of decoherence and relaxation, and in particular for nonlinear open quantum systems (for example, the nonlinear oscillator of Eq.(\ref{qno}) coupled via position to the bath of linear oscillators), there are no known exact solutions. In those cases it is customary to use different approximation methods, such as a perturbative expansion in powers of the interaction Hamiltonian $\hat{H}_{\rm int}$ (Born approximation), and, when applicable, the Markovian approximation (memoryless environment) \cite{gardiner}.  

The model considered in the previous sections (the nonlinear oscillator coupled via density with the bath of linear oscillators) affords an exact solution due to the simplifying property that all terms in the Hamiltonian are diagonal in the joint (system+bath) number basis. 
In order to show how to deal with exact PDEs for observables in generic bosonic open quantum systems that do not have exact solutions, we will now solve  
Eq. (\ref{diff}) for this model considered  in a perturbative expansion in powers of $\hat{H}_{\rm int}$, and compare the results with the exact solution previously found.
The study of other nonlinear models will be left for a future publication.

The perturbative master equation for the reduced density matrix of the quantum nonlinear oscillator is,
to second order in $\hat{H}_{\rm int}$,
\begin{equation}
\frac{d}{dt}  \rho = -i [ (\omega + \delta \omega) \hat{n} + \mu \hbar \hat{n}^2, \rho ] +
t \gamma ( 2 \hat{n} \rho \hat{n} - \hat{n}^2 \rho - \rho \hat{n}^2 ) ,
\nonumber
\end{equation}
where $\hat{n} = \hat{a}^{\dagger} \hat{a}$ is the number operator for the system.
The first term is the free unitary evolution with an environment-renormalized frequency
$\delta \omega = (1/\hbar) \sum_j g_j \langle \hat{n}_j \rangle$, that arises from first order
perturbation theory. The second term is of Lindblad form (but with a time-dependent coefficient),
it arises from second order perturbation theory, and it is responsible for decoherence.
The coefficient $\gamma$  is given by
$\gamma = (1/\hbar^2) \sum_j g_j^2 [ \langle \hat{n}_j^2 \rangle - \langle \hat{n}_j \rangle^2 ]$,
where $\hat{n}_j$ is the number operator for the $j$th oscillator in the environment. The perturbative treatment in powers of $\hat{H}_{\rm int}$ is in fact a perturbative treatment valid for short times  
$ (t / \hbar)   \langle \hat{n}  \rangle \sum_j g_j \langle \hat{n}_j \rangle \ll 1$.
The solution of this approximate master equation  can be straightforwardly found in the Fock basis, and
from there one can evaluate the dynamics for the coherent state amplitude
\begin{equation}
\langle \hat{a}(t)  \rangle = e^{- \frac{\gamma t^2}{2}} \; e^{-i \delta \omega t} \; \alpha(t).
\label{shorttime}
\end{equation}
We see that the coherent state amplitude has an initial quadratic time decay, typical of
quantum systems subjected to perturbations. As expected, this perturbative solution is the same
as that obtained from the exact solution presented in Eq. (\ref{me1}). Indeed, for short times the modulus and amplitude of each decoherence factor $R_j(t)$ can be written as
\begin{eqnarray}
| R_j(t) |  &\approx& 1 - \frac{g_j^2 t^2}{2 \hbar^2}
[ \langle \hat{n}_j^2 \rangle - \langle \hat{n}_j \rangle^2 ] , \nonumber \\
\varphi_j &\approx& - \frac{g_j t}{\hbar} \langle \hat{n}_j \rangle .
\end{eqnarray}
Therefore, the total decoherence coefficient is given by
$R(t) \approx [ 1 - \gamma t^2/2 ] e^{-i \delta \omega t} \approx e^{ - \frac{\gamma t^2}{2}} e^{-i \delta \omega t}$, that coincides with the solution of the perturbative master equation.

We now turn to find the perturbative solution to the exact PDE for observable values, Eq.(\ref{diff}).
Inspired in the solution above, we again propose a solution of the form $f(t) = f_{\alpha}(t) \; f_{\beta}(t)$,
with $f_{\alpha}(t)= \alpha(t)$.
Let us take a perturbative expansion of $f_{\beta}(t)$ in powers of the small parameter
$\epsilon$ of the form
\begin{equation}
f_{\beta} =  f_{\beta}^{(0)} + f_{\beta}^{(1)}+ \ldots,
\end{equation}
where $f_{\beta}^{(0)}$ is independent of  $\epsilon$,  $f_{\beta}^{(1)}$ is linear in $\epsilon$, etc.
Given the initial condition $f_{\beta}(t=0)$=1, then $f_{\beta}^{(0)} =1$. The first order equation  is
\begin{equation}
\dot{f}_{\beta}^{(1)} = f_{\alpha}^{-1} \hat{K}_{\rm int} f_{\alpha} f_{\beta}^{(0)},
\end{equation}
whose solution reads $f_{\beta}^{(1)} = -i t  \sum_j g_j |\beta_j|^2$. Therefore, to first order we obtain
\begin{equation}
f_{\beta}(t) \approx 1 - \frac{i t}{\hbar} \sum_j g_j | \beta_j|^2 \approx e^{- \frac{i t}{\hbar} \sum_j g_j |\beta_j|^2} .
\end{equation}
Integrating upon the probability distributions $P(\beta_j^*,\beta_j)$ we recover the short-time solution
Eq. (\ref{shorttime}),
\begin{eqnarray}
f_S(t) &=& \alpha(t) \times \int \prod_j d^2\beta_j P(\beta_j^*,\beta_j) f_{\beta}(t) \nonumber \\
& \approx & e^{- \frac{\gamma t^2}{2}} e^{-i \delta \omega t} \alpha(t) .
\end{eqnarray}

Another way of obtaining the same result is to use concepts of probability theory. This may turn
out to be useful in other models of decoherence for which long-time solutions for the reduced
dynamics of the system are available \cite{fercook}. Let us call $x_j \equiv g_j |\beta_j|^2$
the stochastic  variable that takes the values $x_j = g_j  n_j$ with probability
$P(n_j) = (\pi \bar{n}_j)^{-1} e^{- n_j / \bar{n}_j}$, where $\bar{n}_j$ is given by the Bose distribution. The mean
value of $x_j$ is $a_j \equiv g_j \bar{n}  = g_j \langle x_j \rangle $, and its variance is
$b_j^2 \equiv \langle x_j^2 \rangle - \langle x_j \rangle^2 = g_j^2 [ \langle \hat{n}_j^2 \rangle -
\langle \hat{n}_j \rangle^2 ]$. The perturbative solution $f_{\beta}(t)$ can then  be written in terms
of these stochastic variables as
\begin{equation}
f_{\beta}(t) = e^{- \frac{ i t}{\hbar} \sum_j x_j} .
\end{equation}
The stochastic variables $x_j$ can be considered as independent and identically distributed.  Therefore,
the stochastic variable $Y \equiv \sum_{j=1} ^N x_j$ belongs to the class of the so-called infinitely divisible  distributions \cite{gnedenko,breiman,reichl}. The behavior of $Y$ depends on whether
the cumulative variance $B_N^2  = \sum_j b_j^2$ is finite or not. In the limit $N \rightarrow \infty$,
$B_N^2$ is finite (central limit theorem), and the probability for $Y$ is Gaussian
\begin{equation}
P(y) = \frac{1}{\sqrt{2 \pi B_N^2}} \exp\left[ - \frac{ (y-\bar{y})^2}{2 B_N^2} \right] ,
\end{equation}
where $\bar{y} = \sum_j \bar{x}_j  = \sum_j g_j \langle n_j \rangle$.  To obtain the reduced dynamics
for the system we need to integrate $f_{\beta}(t)$ over the environmental variables $\beta_j$ weighted
with their probability distributions. This  is equivalent to integrating $e^{-i t y / \hbar}$ over its
probability distribution $P(y)$,
\begin{equation}
\int_{-\infty}^{+ \infty} dy \; e^{- \frac{ i}{\hbar}  t  y } \; P(y) = e^{- \frac{i}{\hbar} \bar{y} t} \;
e^{- \frac{B_N^2 t^2}{2 \hbar^2}} .
\end{equation}
Replacing the expression for $\bar{y}$ and the cumulative variance $B_N^2$, we obtain
our final expression for the reduced dynamics for the coherent amplitude of the system
\begin{equation}
f_S(t)  = e^{- \frac{ \gamma t^2}{2}} \; e^{-i \delta \omega t} \; \alpha(t) ,
\end{equation}
which, again, coincides with the perturbative solution Eq. (\ref{shorttime}).

For other models of decoherence, such as high temperature quantum Brownian motion,
one can proceed along similar lines, {\it i.e.} solve the exact PDE for observables in a perturbative
expansion in powers of the interaction Hamiltonian $\hat{H}_{\rm int}$. It is possible
to introduce an infinitely divisible distribution $Y$ whose probability distribution is not Gaussian,
but given by a Levy distribution, that leads to different time dependencies of the decoherence
factor. For example, for a Lorentzian probability distribution one obtains an exponential decay
\cite{gnedenko,breiman,reichl}.


\section{Conclusions}

In this paper we have generalized the method of exact partial differential equations for observable
values of bosonic systems \cite{bermanbook,vishik}, based on a coherent state representation, to the case when the system
interacts with a bosonic environment. Our method requires to solve, either exactly or approximately,
a PDE containing coupled coherent state degrees of freedom of the system and the environment,
and then to integrate (trace) over the environmental coherent states weighted by their respective
probability distributions. We have exemplified the method with a model of a nonlinear oscillator
interacting via density with a bath of linear oscillators. The simplicity of this model, based on the
fact that all term in the Hamiltonian are simultaneously diagonal in the number basis, allows for an exact solution. 
We demonstrated that the dynamical behavior obtained from the coherent state representation coincides
with that obtained from the reduced density matrix approach. Further development of our method is
required in order to treat other more complicated decoherence models, such as a nonlinear
oscillator coupled through position to the environment. This will involve the study of consistent
perturbative solutions to the exact PDE for observables.

\section{Acknowledgments}

D.A.R.D. thanks F.M. Cucchietti for fruitful discussions. This work was
supported by the Department of Energy (DOE) under Contract No.
W-7405-ENG-36, by the Defense Advanced Research Projects Agency
(DARPA), and by the National Security Agency (NSA).

\end{document}